\documentclass{amsart}

\usepackage{url}

\newcommand{\pplacer}{pplacer}
\newcommand{\EPA}{EPA}
\newcommand{\distallength}{distal\_length}
\newcommand{\edgenum}{edge\_num}
\newcommand{\likelihood}{likelihood}
\newcommand{\likeweightratio}{like\_weight\_ratio}
\newcommand{\marginalprob}{marginal\_prob}
\newcommand{\pendantlength}{pendant\_length}
\newcommand{\postprob}{post\_prob}
\newcommand{\jplace}{\texttt{.jplace}}

\let\oldmarginpar\marginpar
\renewcommand\marginpar[1]{\textit{(see note)}\-\oldmarginpar[\raggedleft\footnotesize #1]%
{\raggedright\footnotesize #1}}

\begin{document}

\title{A format for phylogenetic placements}

\author{Frederick A Matsen}
\author{Noah G Hoffman}
\author{Aaron Gallagher}
\author{Alexandros Stamatakis}

\begin{abstract}
We have developed a unified format for \emph{phylogenetic placements}, that is,
mappings of environmental sequence data (e.g. short reads) into a
phylogenetic tree. We are motivated to do so by the growing number of
tools for computing and post-processing phylogenetic placements, and the
lack of an established standard for storing them. The format is
lightweight, versatile, extensible, and is based on the JSON format which can be
parsed by most modern programming languages. Our format is already
implemented in several tools for computing and post-processing
parsimony- and likelihood-based phylogenetic placements, and has worked
well in practice. We believe that establishing a standard format for
analyzing read placements at this early stage will lead to a more efficient development of
powerful and portable post-analysis tools for the growing
applications of phylogenetic placement.
\end{abstract}

\maketitle

\section{Introduction}

``Phylogenetic placement'' has become popular in the last several years as a
way to gain an evolutionary understanding of a large collection of sequences.
The input to a phylogenetic placement algorithm consists of a reference tree, a
corresponding reference multiple sequence alignment, and a collection of query sequences.
The output of a phylogenetic placement algorithm is a set of assignments of the query
sequences to branches of the tree; there is at least one such assignment for each query.
A query can be assigned to more than one branch on the reference tree to express
placement uncertainty for that query sequence.

Phylogenetic placement methods circumvent several problems associated with
applying traditional phylogenetic algorithms to large,
environmentally-derived sequence data.
The computational burden is decreased compared to constructing a tree containing reference and query sequences \emph{de novo}, resulting in algorithms that can place thousands to tens of
thousands of query sequences per hour {\em and} per processing core
into a reference phylogeny with one thousand taxa. Because computation is performed on each query
sequence individually and independently, placement algorithms are also straightforward to parallelize.
The relationships between the query sequences are not investigated. Hence, the size of the search space
is reduced from an exponential to just a linear number of phylogenetic hypotheses. Moreover,
short and/or non-overlapping query sequences pose less of a problem, as query
sequences are compared to the full-length reference sequences.
Visualization of samples and comparison between samples are facilitated
by the assumption of a fixed reference tree, that can be drawn in a way which
highlights the location and distribution of reads.

The advent of high-throughput sequencing has motivated a growing
interest in phylogenetic placement. The basic idea is as old as
computational phylogenetics \cite{felsensteinML81, klugeFarris69}
although these insertions historically have been considered as just the first step
towards full \emph{de novo} tree reconstruction. Recent implementations have
focused on algorithms for likelihood-based placement, such as
\cite{monierEaLargeViruses08,vonMeringEaQuantitative08}, with more
efficient recent implementations
\cite{berger2011performance,matsen2010pplacer,stark2010mltreemap}. These
tools are being incorporated into popular workflows for microbial
ecology, such as QIIME \cite{caporaso2010qiime} and the next version of
AMPHORA \cite{wu2008simple}. Comparative methods are being developed and
implemented in software \cite{evans2010phylogenetic,matsen2011edge}, and
work is underway to extend a tree viewer
\cite{pirrung2011topiaryexplorer} to visualize placements.
Dedicated algorithms to align reads with respect to reference alignments for subsequent phylogenetic placement are also
being developed \cite{berger2011aligning,mirarabsepp}.

Because of this expansion of activity, standards are needed. The
original versions of \pplacer~\cite{matsen2010pplacer} and
\EPA~\cite{berger2011performance} each implemented their own idiosyncratic
tabular file formats. These ad-hoc formats kept post-analysis tools from being
interoperable and hindered tool comparison.

In this letter, we describe a lightweight file format that will ensure
consistency between tools. Because it adopts JSON (JavaScript Object Notation) \cite{json},
a widely used data interchange standard, and extends the widely used Newick format for
phylogenetic trees, it is straight-forward to parse using existing tools.
It can be used with likelihood, posterior probability, and parsimony-based
placements, can associate an arbitrary number of sequence names associated
with a placement, and can store a generalization of a name list called a \emph{named multiplicity} as
described below. Basic operations such as subsetting arbitrary collections
of placements and merging these lists are easily done. The format can be extended to
incorporate additional information, such as taxonomic assignments.

Although we have made our best efforts to ensure that the format is sufficiently extensible without changing the specification, it may be necessary to change it in the future.
For that reason, the authoritative version of the file format will be maintained on the \url{http://arxiv.org/} server as an online preprint of the same name.
The version described in this document is version 3 of the file format.

\section{Concepts}

We first establish terminology in order to describe the placement
format. As described above, phylogenetic placement is performed by
inserting a collection of query sequences onto a fixed reference tree in
order to optimize a given criterion.
Specifically, for a given set of query sequences the objective is to find an attachment
of each query sequence to the tree that maximizes likelihood or minimizes the parsimony score
for the the reference tree with that (and only that) query sequence attached. Because each
query sequence is placed individually on the tree, the run time complexity is
of order
the product of the number of reference sequences, the number of query
sequences, and the number of sites in the alignment.

There may be more than one good or likely location for a query sequence, and it is
important to record this uncertainty. Uncertainty may be expressed in terms of
placement locations that have equal parsimony scores, in terms of
\emph{likelihood weight ratio} (the ratio of likelihoods of the various
placements), or in terms of posterior probability. Because a given query
sequence thus can be considered to have a collection of placements with varying
certainties, we use the word \emph{pquery} for ``placed query'' to denote the
collection of placements of a single query sequence.

It is also common to obtain several identical sequence reads from deep-sequencing studies.
Furthermore, closely related sequences may exhibit such similar placement results
that a user may wish to group them together for ease of analysis. For this reason, we
allow more than one sequence name to be associated with a given pquery.

Users may simply wish to keep the number of sequences associated with a given
pquery instead of the complete collection of names. More generally, they may
wish to simply have a single floating point number, the \emph{multiplicity} associated with a pquery.
This multiplicity may represent a transformed measure of the quantity of sequences
associated with that pquery, analogous to the transforms that are commonly
applied to ecological count data. For that reason, we also allow the
specification of a named multiplicity associated with a pquery in place of a
list of names.

\section{Design}

One possible representation of a collection of placements would be a single
tree with each placement inserted as a pendant branch.
That design is problematic for representations of
uncertainty; if each possible location for every query sequence were
represented as a pendant branch, then it would be difficult to distinguish the
pendant edges that resulted from uncertainty with those resulting from multiple
query sequences. Subsetting collections of placements would require tree
``surgery''. Furthermore, packing everything into a tree would make placement-specific
metadata such as multiple confidence measures difficult to keep track of.
Also, visualizing a reference tree with 1,000 taxa and 10,000 queries {\em and} with several
placements per query may become computationally and visually
cumbersome.

These considerations led us to develop a format where the
placements are represented as a list, and their branch assignments
are indexed by numbered edges of the reference tree. Each placement is
associated with entries for a collection of \emph{fields}, which can contain
arbitrary data about the placement. With such a list-based format,
subsetting pqueries becomes trivial.

With the separation of reference set and placements in mind, our goals
in designing the format were: to adopt a popular extensible open
standard human-readable file format, to ease parsing between languages
and tools, and to deploy a light-weight format that can handle large
collections of placements on large reference trees without requiring too
much space. We chose JSON, since it satisfies all of the above
criteria.

Using the JSON syntax, one option would be to individually associate each
placement with an arbitrary collection of information using key-value pairs for
each placement. However, doing so would have created a substantial file size overhead, as
the total number of characters used to represent the keys would be about the
same as the total number of characters used to represent the data. Because of
this, the field titles are written out only once, and every placement just
supplies the data as an array with entries in the correct corresponding order, as described
below.

\section{Specification}

Files using the format described in this paper will use the
\jplace~suffix, which is short for JSON placement.

The basic types in a JSON file are \verb|Array|, \verb|Boolean|, \verb|Number|, \verb|Object|,
\verb|String|, and \verb|null|. These are familiar terms except \verb|Object|, which is a
list of colon separated key-value pairs, where the keys are strings and
the values are arbitrary types. A JSON file contains a single JSON
object.

In \jplace~files, the fundamental object contains a list of four keys:
``tree'', ``fields'', ``placements'', ``metadata'', and ``version''. We
will describe each of these in succession, but this need not correspond
to their order in the JSON object. Indeed, the order of key-value pairs
in a JSON object is unspecified.

\subsection{tree}

To represent the tree, we extend the well-known Newick file format. In that
format, commas and parentheses are used to display the structure of the tree.
The taxon names (leaf labels) are inserted as plain text. It is also common to label internal nodes
with strings appearing after the closing of a parenthesis. It is also possible
to label edges of the tree with strings enclosed in square brackets. For
example, the tree

\begin{verbatim}
((A:.01[e], B:.01)D:.01[g], C:.01[h]);
\end{verbatim}
is a tree with some edge labels and some node labels.

We extend this format with edge numberings in curly braces:
\begin{verbatim}
((A:.01[e]{0}, B:.01{1})D:.01{3}[g], C:.01{4}[h]){5};
\end{verbatim}
These edge numberings index the edges for the placements.
We use curly braces to distinguish between our edge numberings and
other edge label values such as posterior probability or bootstrap branch (bipartition) support.

Although not required for parsing, we use a convention that placement
algorithms should use a pre-defined edge numbering. Specifically, we enforce
that branches are labeled by a depth-first traversal (descending left subtree
first and starting at the root/top-level node in the reference
input tree) and we assign branch numbers by a post-order traversal.
This strict definition is convenient to ensure one-to-one comparability of
results obtained from various placement algorithms.

We also require the output tree to be identical as a planar tree to the input
reference tree, that is, the subtree ordering and top-level trifurcation must
remain unchanged.
In the case of parsimony-based placements, the reference tree may optionally be
represented without branch lengths.

\subsection{fields}

The value associated with ``fields'' is an array of strings specifying
the headers in the same order as the arrays of placement data. For example, the default
fields for a maximum likelihood EPA or pplacer run are
\edgenum,
\likelihood,
\likeweightratio,
\distallength, and
\pendantlength.

The \edgenum~specifies the placement edge, and is necessary for all
placements. The \pendantlength~is the branch length for the
placement edge, and \distallength~is the length from the distal (away
from the root) side of the reference tree edge to the placement
attachment location. The \likelihood~is the likelihood of the tree with
the placement attached, which may be calculated from an alignment with
columns masked out that do not appear
in the read. For that reason, the log likelihood of the placement may be better (closer to zero) than the log likelihood of the reference tree on the full-length alignment.
The \likeweightratio~is the ratio of that placement's
likelihood to that of the other alternate placements for that read. For a pplacer
posterior probability run, the marginal likelihood
\marginalprob~and the posterior probability \postprob~are also
included.

In contrast to pplacer, EPA optimizes three branch lengths associated with
a placement: the pendant branch length, the distal branch length, and the
proximal branch length.
Thus, the EPA output could be extended to comprise the full information
generated by the EPA algorithm by adding a proximal\_length field.
Because the currently available downstream placement analysis tools (e.g.,
guppy) do not use this additional information, it is not
included in the EPA \jplace\ output file at present.

The corresponding fields for parsimony-based placements (currently only available in EPA) are
\edgenum\ and parsimony.
The parsimony field just contains the parsimony score of the placement as an integer.

\subsection{placements}

The value associated with the ``placements'' key is the list of
placements grouped into pqueries. The representation of each pquery is a
JSON object of its own, with two keys: ``p'', for placements,
and either ``n'' for names or ``nm'' for names with multiplicity.
The value associated with ``p'' is the list of placements for that pquery with entries corresponding to the fields in the order set up by the ``fields'' described above.
The list of placements shows possible placement locations along with their confidence scores and other information.
The value associated with ``n'' is a list of names associated with that pquery.
Although an arbitrary list of names can be associated with a pquery, the typical use will be to collect placement information for identical or closely related sequences.
The value associated with ``nm'' is a list of \emph{named multiplicities}, which as simply ordered pairs of a name and then a positive floating point value.
As described above, multiplicity can be used to keep track of the number of sequences associated with that name or a transform thereof.

For parsimony-based placements we require all equally parsimonious placements of a
query to be included in the output file.
This is to enable easy comparison between parsimony-based placement methods; if only one of the best-scoring placements is arbitrarily
selected in one way or another, comparing programs based on our standard will become error-prone and biased.

\subsection{Other keys}

There are also two other keys in the fundamental JSON object. The first,
``version'', is mandatory, and indicates an integer version number of
the format. The version described in this paper is 3. The second,
``metadata'', is optional and keys an arbitrary object for metadata. It can describe how the
placement file was generated, which phylogenetic model was used, and so on.
In EPA and pplacer we include the full command line string of the placement program invocation
to allow for easy reproducability of results.

\subsection{A small example}

\begin{verbatim}
{
  "tree": "((A:0.2{0},B:0.09{1}):0.7{2},C:0.5{3}){4};",
  "placements":
  [
    {"p":
      [[1, -2578.16, 0.777385, 0.004132, 0.0006],
       [0, -2580.15, 0.107065, 0.000009, 0.0153]
      ],
     "n": ["fragment1", "fragment2"]
    },
    {"p": [[2, -2576.46, 1.0, 0.003555, 0.000006]],
     "nm": [["fragment3", 1.5], ["fragment4", 2]]}
  ],
  "metadata":
  {"invocation":
    "pplacer -c tiny.refpkg frags.fasta"
  },
  "version": 3,
  "fields":
  ["edge_num", "likelihood", "like_weight_ratio",
               "distal_length", "pendant_length"]
}
\end{verbatim}

\subsection{Tabular representation}

The JSON object can be readily transformed into a tabular format to
more easily summarize or explore the data using statistical tools or a
relational database. With the addition of an index (``placement\_id'')
to form a relation between placements and sequence names, two tables
are sufficient: one with columns ``placement\_id'' followed by each of
the fields contained by each pquery array, and another providing a
mapping of every ``placement\_id'' to each of the corresponding
sequence names or named multiplicities.  This transformation can be performed efficiently
using any modern high level language with a JSON parsing library. Such a
representation of the data is useful for supporting analyses that
involve grouping and partitioning placements and sequences.

\section{Tools}

The latest versions of EPA (\url{http://github.com/stamatak/standard-RAxML}) and pplacer (\url{http://matsen.fhcrc.org/pplacer/}) both produce these files.
The guppy program in the
pplacer suite has a number of subcommands that allow
transformations and filterings of these files (manuscript in preparation).
MePal, an implementation of placement using an alignment-free generalization to
indels of Felsenstein's phylogenetic pruning algorithm
\cite{westesson2011alignment}, now imports and writes out this format as well.
The TopiaryExplorer \cite{pirrung2011topiaryexplorer} tree visualization package
is now in the process of being extended to read this format for visualization.

\section{Conclusion}

We have designed a unified format for phylogenetic placements.
The format is lightweight,
flexible, and is based on JSON, a well-established data interchange standard.
The format handles placement uncertainty and allows for multiple sequence
names to be associated with the placement of a single sequence. Current
versions of two placement software packages have already adopted the
format, and others are in the process of doing so.

\bibliography{placefmt}
\bibliographystyle{plain}
\end{document}